\documentclass[pdflatex,sn-basic]{sn-jnl}

\jyear{2022}%

\theoremstyle{thmstyleone}%
\theoremstyle{thmstyletwo}%

\theoremstyle{thmstylethree}%

\begin{document}
	
	\title[3D PTV System for Atmospheric Flows]{Imaging-based 3D Particle Tracking System for Field Characterization of Particle Dynamics in Atmospheric Flows}
	
	
	\author[1,2]{\fnm{Nathaniel} \sur{Bristow (ORCID: 0000-0001-7287-7635)}}
	
	\author[1]{\fnm{Jiaqi} \sur{Li (ORCID: 0000-0002-1201-7489)}}
	
	\author[3]{\fnm{Peter} \sur{Hartford}}
	
	\author[4]{\fnm{Michele} \sur{Guala (ORCID: 0000-0002-9788-8119)}}
	
	\author*[1,2]{\fnm{Jiarong} \sur{Hong (ORCID: 0000-0001-7860-2181)}}\email{jhong@umn.edu}
	
	\affil[1]{\orgdiv{Mechanical Engineering}, \orgname{University of Minnesota -- Twin Cities}, \orgaddress{\street{111 Church Street SE}, \city{Minneapolis}, \postcode{55455}, \state{Minnesota}, \country{USA}}}

	\affil[2]{\orgdiv{St Anthony Falls Laboratory}, \orgname{University of Minnesota -- Twin Cities}, \orgaddress{\street{2 3rd Ave SE}, \city{Minneapolis}, \postcode{55414}, \state{Minnesota}, \country{USA}}}
	
	\affil[3]{\orgdiv{Aerospace Engineering and Mechanics}, \orgname{University of Minnesota -- Twin Cities}, \orgaddress{\street{110 Union St SE}, \city{Minneapolis}, \postcode{55455}, \state{Minnesota}, \country{USA}}}
	
	\affil[4]{\orgdiv{Civil, Environmental, and Geo- Engineering}, \orgname{University of Minnesota -- Twin Cities}, \orgaddress{\street{500 Pillsbury Drive SE}, \city{Minneapolis}, \postcode{55455}, \state{Minnesota}, \country{USA}}}

	
	\abstract{A particle tracking velocimetry apparatus is presented that is capable of measuring three-dimensional particle trajectories across large volumes, of the order of several meters, during natural snowfall events. Field experiments, aimed at understanding snow settling kinematics in atmospheric flows, were conducted during the 2021/2022 winter season using this apparatus, from which we show preliminary results. An overview of the methodology, wherein we use a UAV-based calibration method, is provided, and analysis is conducted of a select dataset to demonstrate the capabilities of the system for studying inertial particle dynamics in atmospheric flows. A modular camera array is used, designed specifically for handling the challenges of field deployment during snowfall. This imaging system is calibrated using synchronized imaging of a UAV-carried target to enable  measurements centered 10\,m above the ground within approximately a 4\,m$\times$4\,m$\times$6\,m volume. Using the measured Lagrangian particle tracks we present data concerning 3D trajectory curvature and acceleration statistics, as well as clustering behavior using Vorono\"{i} analysis. The limitations, as well as potential future developments, of such a system are discussed in the context of applications with other inertial particles.}

	\keywords{3D particle tracking, atmospheric flows, inertial particles}
	
	
	
	\maketitle
	
	\section{Introduction}\label{sec:intro}
	\normalsize
	A wide variety of geophysical processes involve atmospheric transport of inertial particles, such as wind-blown sand and dust, sea sprays, and snowfall, wherein the details of particle kinematics can have long-ranging implications for society in terms of the environment, climate, and weather. Aeolian transport of sand drives major changes in landscape, not only on Earth but also on other planets \citep{lapotre2016large}, and the related suspension of mineral dust aerosols in the atmosphere, and their subsequent settling, may greatly influence global climate \citep{kok2018global}. For example, sea sprays, which can broadly also be characterized in the same manner as inertial particles in atmospheric turbulence, play an important role in the momentum and thermal fluxes at the air-sea interface, and thus must be modeled appropriately for the forecasting of tropical cyclones \citep{emanuel2018100}. Snow particles, which are the primary focus of the study herein, are particularly complex inertial particles due to their intricate morphologies, causing different tumbling behaviors and turbulence interactions that modulate their fall-speed and thereby the resultant ground accumulation heterogeneity \citep{garrett2014observed,JLi2021}. For such reasons, it is critical to better understand the kinematics of atmospheric inertial particles and their interaction with the surrounding flow. 
	
	Such atmospheric flow, however, is naturally characterized by a broad range of turbulence scales, where Reynolds numbers are typically on the order of $10^6$. The largest motions that drive bulk transport can be captured by remote sensing techniques, such as time-of-flight light detection and ranging (LiDAR) from satellites \citep{huang2015detection}, for tracking aerosol movements, or Doppler LiDAR \citep{lundquist2017assessing} for wind measurements in the atmospheric boundary layer (ABL). However, resolution of the small-scale motions that govern the dynamics of individual particles, and thus influence important parameters for modeling such as fall-speed, requires alternative approaches, such as imaging, for particle-level tracking. 
	
	Imaging systems for particle tracking and velocimetry are commonplace in the laboratory setting, implemented in both 2D and 3D, the latter with multi-camera arrays (for an overview, see \citep{discetti2018volumetric}). The 3D approach is more challenging due to the need for precise calibration of the camera system to be able to triangulate or otherwise reconstruct particle positions accurately, but also due to the difficulty of subsequently linking these particles across time steps to form long Lagrangian trajectories that are needed to study inertial particle kinematics. Great advancements have been made in both respects, such as with iterative particle reconstruction (IPR; \citep{wieneke2012iterative}) and predictive tracking (e.g., Shake-the-Box; \citep{schanz2016shake,novara2019multi}), both of which aid in enabling long Lagrangian trajectories to be obtained from dense particle images. However, such methods have generally been limited to measurement domains of $\sim$1\,m \citep{schanz2016towards,terra2019aerodynamic}, due to combined challenges of tracer seeding, illumination, and calibration, and thus have seen very limited implementation in the field due to technical challenges. 
	
	That being said, notable examples of particle imaging-based flow measurement have been conducted in the field. These include particle image velocimetry (PIV) or particle tracking velocimetry (PTV) in the atmospheric surface layer (ASL) for both wind energy research \citep{hong2014natural,abraham2020dynamic,wei2021near} and boundary layer turbulence studies \citep{hommema2003packet,morris2007near,toloui2014measurement,heisel2018spatial}, PTV for snow settling in the ASL \citep{Nemes2017,CLi2021,JLi2021}, as well as water surface velocimetry \citep{perks2020towards}. In the case of the ASL, both artificial tracers (smoke, fog) and natural tracers (snow) have been used, while in the latter case for water surface velocimetry the patterns of surface reflections from the water itself may be used, or else artificial tracers introduced. However, except for \citep{wei2021near}, all the cited approaches are two-dimensional, either in a light sheet or on a surface. 
	
	For the case of snow settling, such 2D studies have provided valuable insights into the enhancement of settling velocities due to turbulence \citep{Nemes2017}, clustering behavior influencing settling \citep{CLi2021}, and preferential sampling of sweeping motions from vortices \citep{JLi2021}. Nevertheless, three-dimensional measurements of snow particle motions are necessary in order to fully resolve their kinematics. In particular, both trajectory curvature and particle clustering may be biased on resolving only in 2D, and therefore 3D measurements of snow particle transport are needed.
	
	Adaptation of three-dimensional particle tracking to the field is even rarer, compared to cases mentioned for 2D measurement, due to limitations from calibration procedures, particle seeding, and illumination. A recent noteworthy implementation of 3D PTV in the field, however, is given by \cite{wei2021near}, who used a multi-camera system to measure the time-averaged flow structure and vorticity in the wake of a full-scale vertical axis wind turbine. They achieved measurements in a large volume (10\,m$\times$7\,m$\times$7\,m), though with a low particle concentration, as they had to use artificial snow as tracers introduced upstream. Multiple runs had to be performed and datapoints spatially binned over the ensemble to obtain a single time-averaged flow field, as the entire domain could not be sampled simultaneously. Thus, turbulent fluctuations could not be captured. 
	
	Other related work focused on obtaining Lagrangian trajectories in large-scale measurements has been conducted in the field of bio-locomotion \citep[e.g.,][]{theriault2014protocol,muller2020calibration}, tracking the motions of birds and of fish. These studies generally involve sparse objects, as usually only a few points of interest within the field of view are to be tracked, but have involved innovative calibration approaches to enable tracking in large volumes. The study herein builds upon these calibration approaches, as will be discussed in \ref{sec:methods:calibration}.
	
	The aim of the present work is to track complex 3D motions of snow particles as they settle in atmospheric turbulence, and in particular to obtain a sufficiently high number of Lagrangian trajectories to enable the estimate of acceleration, curvature, and high order statistics. Here we seek to overcome challenges inherent to the snow settling case, measuring 10\,m above the ground in a large volume while also dealing with high concentrations of particles in camera images. In the following, we describe our methodology for achieving these aims through the development of a field-scale 3D imaging system (\ref{sec:methods:system}--\ref{sec:methods:particletracking}), including preliminary validation experiments (\ref{sec:methods:validation}), after which we show results from a field deployment in April 2022 (\ref{sec:experiment}) to demonstrate the capabilities of this system in terms of the new data and analysis afforded by such measurements (\ref{sec:results}). This work is conducted as part of the Grand-scale Atmospheric Imaging Apparatus project, hereafter referred to as GAIA.
	
	\section{Methods}\label{sec:methods}
	\subsection{System description}\label{sec:methods:system}
	
\begin{figure}[h]
		\centering
		\includegraphics[width=1.0\textwidth]{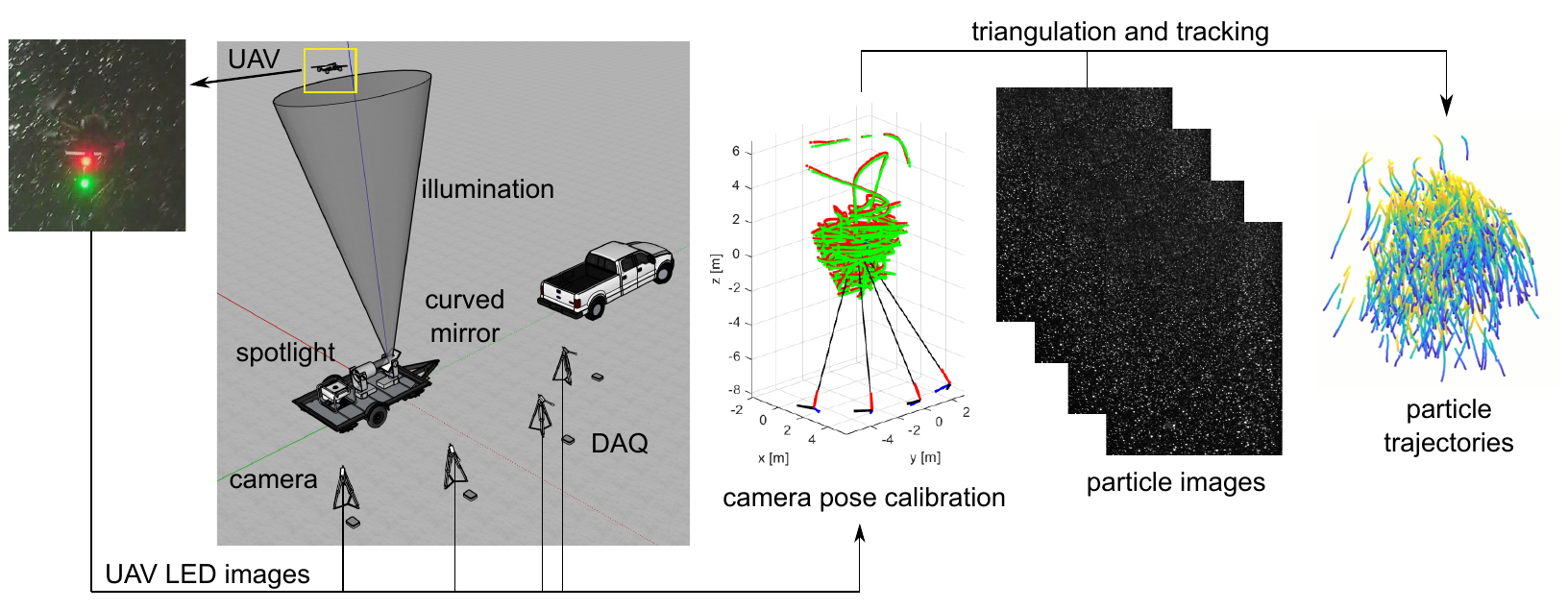}
		\caption{Schematic illustrating overview of GAIA-PTV workflow.}
		\label{fig:methods:schematic}
	\end{figure}

	The principal components of the field-scale 3D imaging system developed under the GAIA project, hereafter referred to as GAIA-PTV, and the associated workflow, are depicted in Fig.~\ref{fig:methods:schematic}. In the following, we provide a brief overview followed by a more detailed description. Firstly, GAIA-PTV involves a modular, multi-camera imaging array that, once calibrated, enables the desired 3D particle triangulation and tracking. To obtain the calibration of the camera array, an unmanned aerial vehicle (UAV), equipped with its own lighting, is flown within the measurement volume while being imaged by all cameras. Images collected during the calibration procedure as well as the particle images taken during the experiment are then processed as will be described in Sections~\ref{sec:methods:calibration} and \ref{sec:methods:particletracking}. 
	This system was designed with a few key criteria in mind. Due to the difficult working conditions in the field during snowfall, GAIA-PTV needed to be easy to use and capable of rapid deployment. Furthermore, it was desirable for deployment configurations to be flexible and enable large distances between cameras. Such flexibility is needed in the field when snow concentration, and thus image density, is outside our control, and instead, we may desire to adjust the particle image density in each camera by changing the camera separation distances. Lastly, we needed to measure snow settling dynamics in a volume of at least several meters in each dimension, centered high enough above the ground to avoid surface effects on the flow.
	
	To meet these demands, GAIA-PTV implements a modular approach. We use a dedicated single-board-computer (SBC) for each camera and wireless communication over a local network, using a COMFAST Outdoor WIFI Antenna. This local network can also be used for wireless camera synchronization for frame rates $<100$ Hz. Thus, each camera is a stand-alone unit that can be moved and set up easily. The only cables between cameras are for synchronization $>100$ Hz, wherein GPIO connectors from separate cameras would be wired together, setting one camera as ``primary'' to output a trigger signal when exposure begins causing the remaining ``secondary'' cameras to simultaneously begin capture. Having dedicated SBC's is also a cost-saving approach, removing the need for a more expensive central processing computer capable of handling the bandwidth from the multiple cameras, and thus enables easy future upgrades of the system with more than 4 cameras. This was achieved using Nvidia Jetson Nano SBC's, each equipped with a wireless antenna for communication and a solid-state drive (SSD) for rapid data storage such that images do not need to be buffered on the camera and can be continuously recorded at a sufficiently high frame rate (e.g., 200 Hz). Nvidia's Jetson platform provides a full Linux operating system with enough I/O for all auxiliary devices, needing only a 5 V, 10 W power supply easily provided from a battery pack. 
	
	The cameras used are FLIR Black Fly S U3-27S5C color units with the Sony IMX429 sensor, pixels measuring 4.5 $\mu$m, capable of 95 fps continuous recording at 2.8 MP (1464$\times$1936\,pix.), or up to 200 fps at 0.7 MP, using decimation. Each camera is paired with a 16\,mm Fujinon lens, with 3.45 $\mu$m pixel pitch and an aspherical lens design minimizing radial distortions.
	
	As stated above, an additional requirement for the system was that it be able to image snow particles in a measurement region high above the ground, approximately 10\,m. This particular choice of height was motivated by the goal of avoiding the flow effects of local surface ``roughness'' elements (e.g., bushes, small topographic changes) on snow settling, ensuring a more uniform light background for the four-camera images, avoiding contamination from re-suspended snow particles, and lastly, to match previous experiments involving 2D tracking that were designed under similar considerations \citep{Nemes2017,CLi2021,JLi2021}. This design criterion introduced two challenges for the system. First, our lighting system to illuminate the snow particles needed to be strong enough for those distances, and second, the camera calibration would need to be done far above the ground.
	
	The sampling volume to be imaged measures typically around 4\,m$\times$4\,m$\times$6\,m ($x$, $y$, $z$). Therefore, to illuminate the snowflakes throughout this large of a volume centered at a height of 10\,m, a 5 kW spotlight is used, similar to that used by \cite{hong2014natural} and \cite{toloui2014measurement}. The beam is expanded at 18 degree divergence half angle such that it spread to a 4\,m wide region in the measurement volume, after reflecting off of a 45 degree polished aluminum mirror. The conical illumination region is then able to be further modified by adjusting the curvature of the mirror such that the volume is elongated in the desired direction (e.g., downstream). 
	
	\subsection{Calibration}\label{sec:methods:calibration}
	
	The camera calibration cannot follow traditional approaches taken in the laboratory, and instead is implemented using a UAV equipped with two rigidly connected lights that is flown throughout the measurement domain. While being imaged by the camera array, the UAV samples the entire measurement volume from edge to edge. The basis of this approach, which follows the method by \cite{theriault2014protocol}, is to move an ``object'' of known dimensions through the measurement volume while identifying image coordinates of the object keypoints in each camera. Here, our object consists of two LEDs, one green and one red, held at a fixed separation of 34.5\,cm by a carbon fiber tube, with the object ``keypoints'' being the LEDs themselves. This apparatus is attached to the UAV, which needs to be flown carefully throughout the measurement domain.
	
	Snow tracking deployments, including those described in \ref{sec:experiment}, are conducted at night, with artificial illumination. In such a case, the extent of the measurement volume in the horizontal plane, and the UAV within it, is visually apparent to the flight operator. Therefore, the UAV flight plan can be implemented manually, keeping within this $x$--$y$ domain while monitoring the UAV altitude to stay within the desired region. A Holybro V500 quadcopter UAV, with 400\,mm wheelbase, is used for this purpose, with a Pixhawk flight controller operating on the open-source ArduPilot platform wherein the Mission Planner software can be used for monitoring the UAV.
	
	However, it should be noted that GAIA-PTV was also designed with the future goal in mind to be capable of other types of 3D PTV measurements in daylight as well, where no artificial lighting would be used, in which case a precise, automated flight plan may need to be charted with GPS coordinates. Mission Planner enables many features for flight planning that are useful in this regard, including tools for automated scanning of regions at various altitudes, as well as the ability to inject real-time kinematic positioning (RTK) corrections to the GPS. The UAV is therefore equipped with an RTK-capable Ublox M8P GPS unit that, upon integration with NTRIPP data stream provided from a local Minnesota Department of Transportation (MNDOT) base station, is capable of providing $\sim$cm scale accuracy to the UAV positioning. Regular GPS units are typically only $\sim$1\,m accurate, at best. 
	
	The UAV flight for calibration typically lasts about 5 minutes while the cameras capture images at 30 fps in order to obtain several thousand positions at which the UAV LEDs are seen by all cameras. All images are processed using Python with in-house code to find all potential green and red LED candidates. A color filter is used in image processing, selecting the green and red bands in the RGB images separately, and an intensity threshold is applied. Poor candidates are further filtered using a temporal smoothness constraint to the LED positions in the 2D images.
	
	To obtain a calibration mapping from the image to the object space, the 8-point algorithm is used \citep{hartley2003multiple} to initially estimate camera pose (i.e., position and orientation), based on a pinhole model. This estimate is then refined using sparse bundle adjustment (SBA) \citep{lourakis2009sba}, implemented in ``easyWand'' by \cite{theriault2014protocol}. The uncertainty of the calibration results that are obtained following the SBA is assessed using two different metrics: reconstructed LED separation in 3D space, varying for each frame captured; and reprojection error, given as the r.m.s. distance between original and reprojected keypoints. If the mean reprojection error are too large, i.e. $>$0.5 pixels, data points with the largest reprojection errors are iteratively removed and the calibration re-calculated. Otherwise, more calibration points could be collected with further flights in order to reduce error. Such calibration flights are typically conducted both before and after particle image datasets are collected.
	
	\subsection{Particle tracking}\label{sec:methods:particletracking}
	
	After collecting particle images and obtaining a camera calibration, Lagrangian trajectories are extracted using predictive 3D PTV following the implementation by \cite{tan2020introducing} with OpenLPT. This open-source implementation of 3D PTV is based on the Shake-the-Box (StB) algorithm by \cite{schanz2016shake}, with additional improvements made for ghost particle rejection. StB has been shown to be a robust predictive tracking approach that iteratively solves the problems of triangulation and matching/linking across frames simultaneously, leveraging the expected particle position information to handle dense particle images (e.g., 0.1 ppp). In OpenLPT, initializing trajectories with which to predict subsequent particle positions requires at least 4 time steps. Following initialization, at each frame a combination of active, inactive, and exited tracks are determined. The active tracks are a combination of short trajectories that have not yet reached the needed initialization length and long tracks that have been successfully predicted. Once an active track is lost, it is added to the inactive tracks list, and if the particle departs the specified measurement domain, it is added to the list of exited tracks. 
	
	Once trajectories of particles are obtained, their paths must be smoothed before computing higher-order quantities such as acceleration and curvature. A Gaussian filter kernel is convolved with the particle position vector in each component, with an optimal kernel size determined following the approach of \citep{Nemes2017}. If the kernel is too small, acceleration data is contaminated by high-frequency noise, whereas if it is excessively large, fine-scale motions may be overly attenuated. In the results presented in \ref{sec:results}, this filter length was 0.26\,s, or $\sim3\tau_{\eta}$, where $\tau_{\eta}$ is the Kolmogorov timescale, which is similar to that used in prior studies \citep{Nemes2017,CLi2021}.
	
	\subsection{Method validation and demonstration}\label{sec:methods:validation}
	
\begin{figure}[h]
		\centering
		\includegraphics[width=0.5\textwidth]{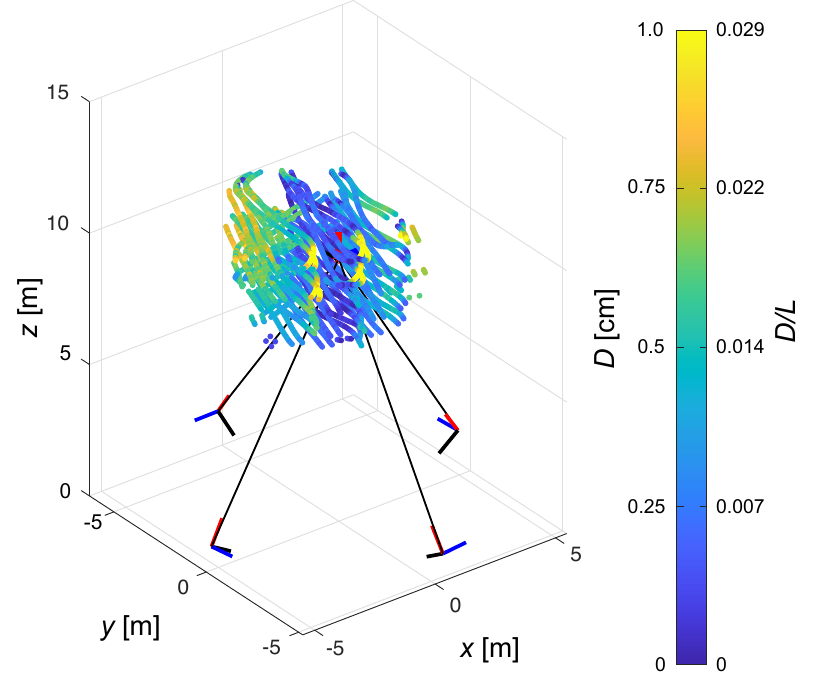}
		\caption{Spatial distribution of distances, $D$, between UAV light pairs after reconstruction, plotted with calibrated camera poses indicated by each set of axes.}
		\label{fig:methods:val:drone}
	\end{figure}

	Prior to the deployment of GAIA-PTV in the snow, two experiments were undertaken to validate the approach. First, the UAV-based calibration method was tested at scale to determine if this methodology could provide sufficiently accurate camera pose estimation. Figure~\ref{fig:methods:val:drone} shows a spatial map of the path taken by the UAV, where markers denote the reconstructed positions of the UAV lights colored by their deviation from the true distance between the lights, $D$. For this case the UAV’s flight was automated to scan throughout a volume measuring approximately 4\,m$\times$4\,m$\times$4\,m  using the equipped RTK GPS unit. It can be seen that reconstructed errors are largest near the edges of the calibration volume, where they are at most $\sim$1\,cm, and much less throughout the core of the volume. A separate scale is added where $D$ is normalized with the measured distance between lights of 34.5 cm, where relative errors are below 3\%. 
	
\begin{figure}[h]
		\centering
		\includegraphics[width=1.0\textwidth]{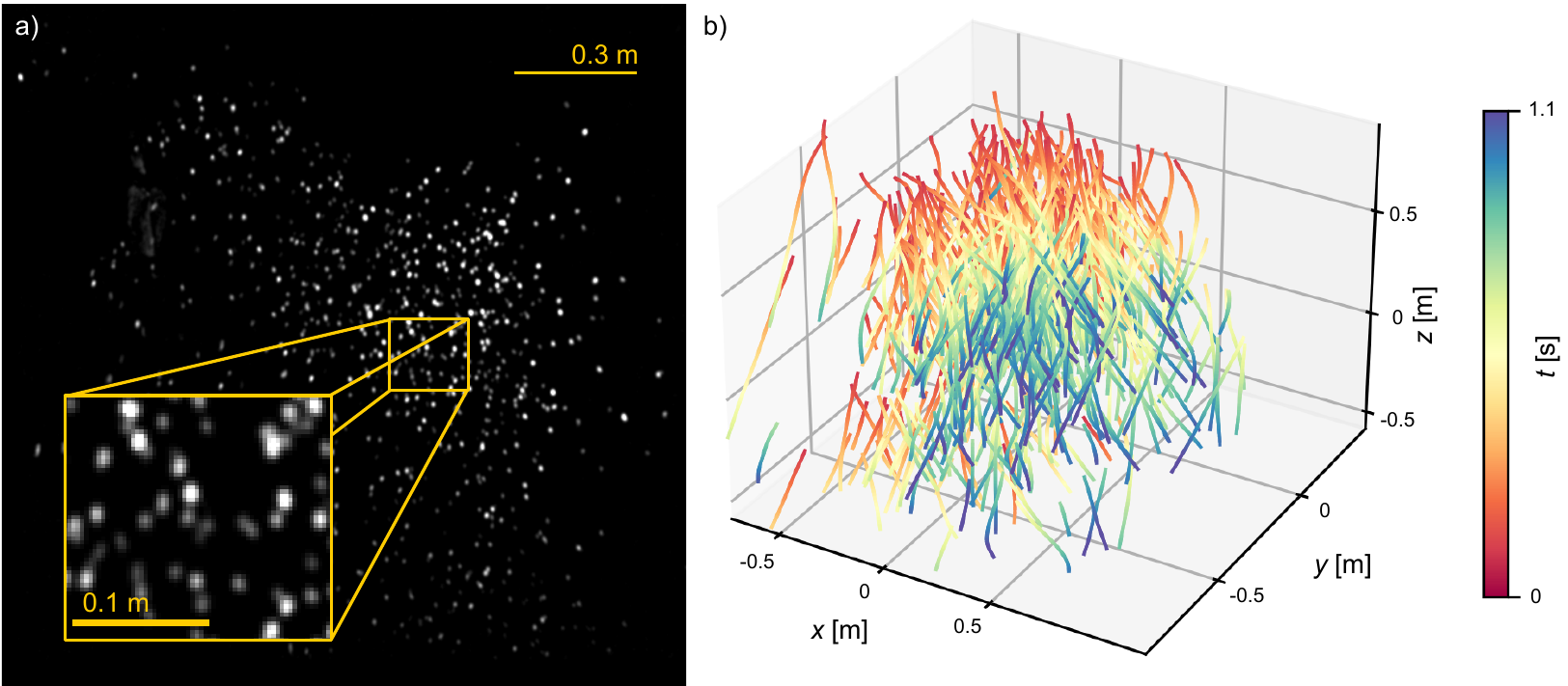}
		\caption{a) A sample confetti image; b) all confetti trajectories with colors indicating time from beginning of image sequence.}
		\label{fig:methods:val:confetti1}
	\end{figure}

	A further experiment was performed under controlled conditions to test the tracking capabilities provided by the calibration when input to the OpenLPT software. For this experiment, standard hole-punch confetti paper particles (6\,mm diameter) were dropped manually in still-air conditions in a large warehouse environment. Though it was not feasible to densely fill an entire measurement with such particles, they mimicked many of the conditions expected to be encountered in the field while also enabling some ground-truthing of their fall-speed. In particular, the disk-shaped confetti were expected to exhibit more complex trajectories with tumbling motions similar to that can be exhibited by snowflakes, such as those with dendrite morphologies. Cameras were positioned in a fan array at 11\,m radial distance from the center of the measurement volume in order to reproduce a similar scale to be measured over in the field. The calibration here was not performed with a UAV, as the experiment was indoors, and instead the LEDs and carbon fiber rod were held by hand and moved through the measurement volume manually. A sample confetti image (after enhancement) and resultant trajectories are shown in Fig~\ref{fig:methods:val:confetti1}, where it can be seen that long trajectories are able to be captured. Furthermore, it was estimated that approximately 93\% of confetti particles could be tracked, based on the identified particles in each image (after filtering out double-counted particles whose centroids were within 1 pixel of each other) and the sum of the durations of all particle tracks (in number of frames), which gave an estimate of the average particles tracked per frame. 
	
\begin{figure}[h]
		\centering
		\includegraphics[width=1.0\textwidth]{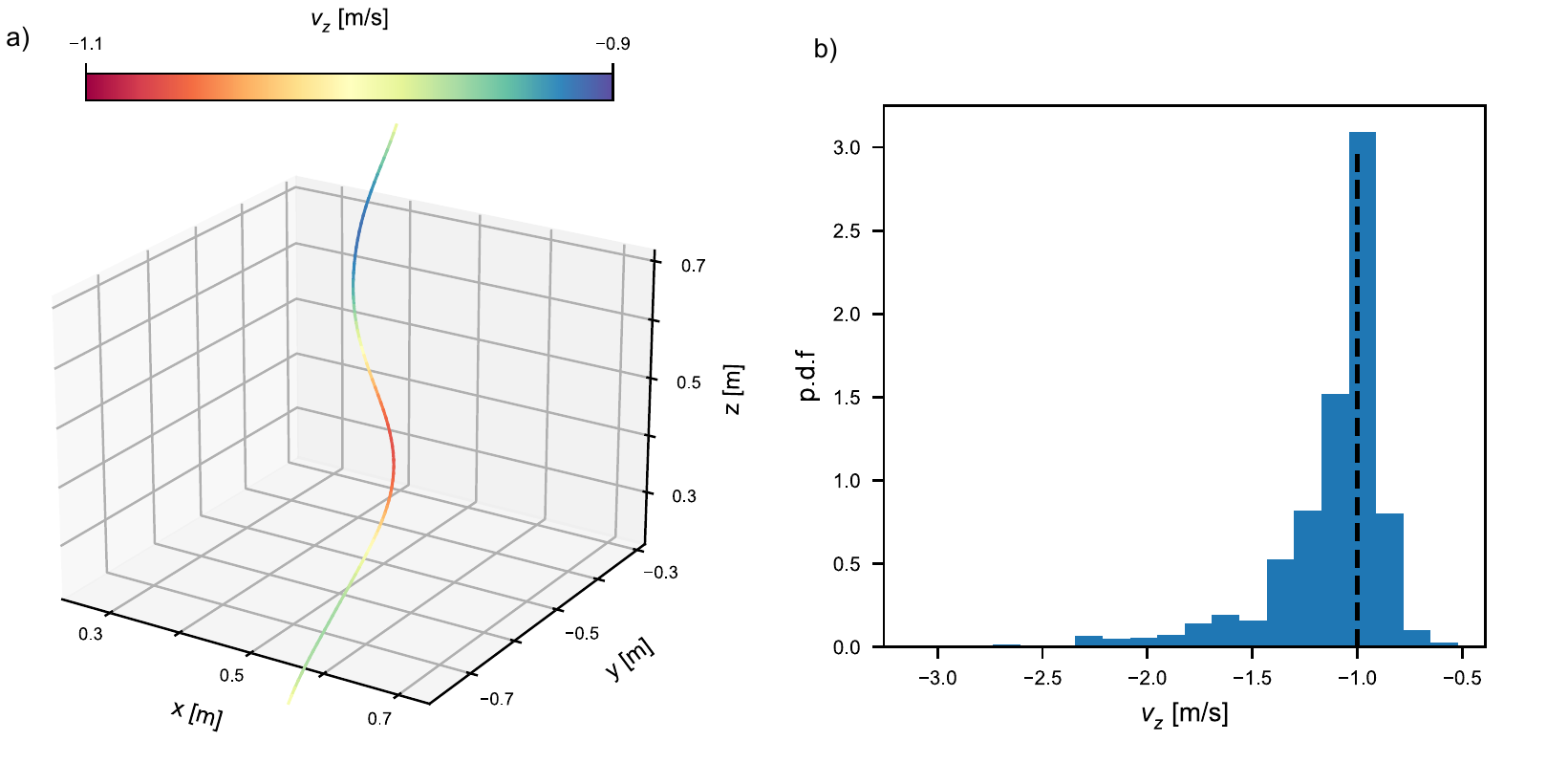}
		\caption{a) Sample confetti trajectory; b) Probability density function of mean fall speeds for individual particles, with dashed vertical line indicating separately measured average fall speed.}
		\label{fig:methods:val:confetti2}
	\end{figure}

	Looking closer at individual trajectories, as in Fig.~\ref{fig:methods:val:confetti2}a, it can be seen that the fine-scale motions of the confetti disks can be resolved, with an oscillation in the trajectory as the disk tumbles end-over-end. The average fall-speed of individual confetti particles was also measured in independent experiments and compared with the distribution of particle fall-speed measured with PTV, which show close agreement (Fig.~\ref{fig:methods:val:confetti2}b). 
	
	\section{Experiment}\label{sec:experiment}
	
	\subsection{Field site}\label{sec:experiment:fieldsite}
	
\begin{figure}[h]
		\centering
		\includegraphics[width=1.0\textwidth]{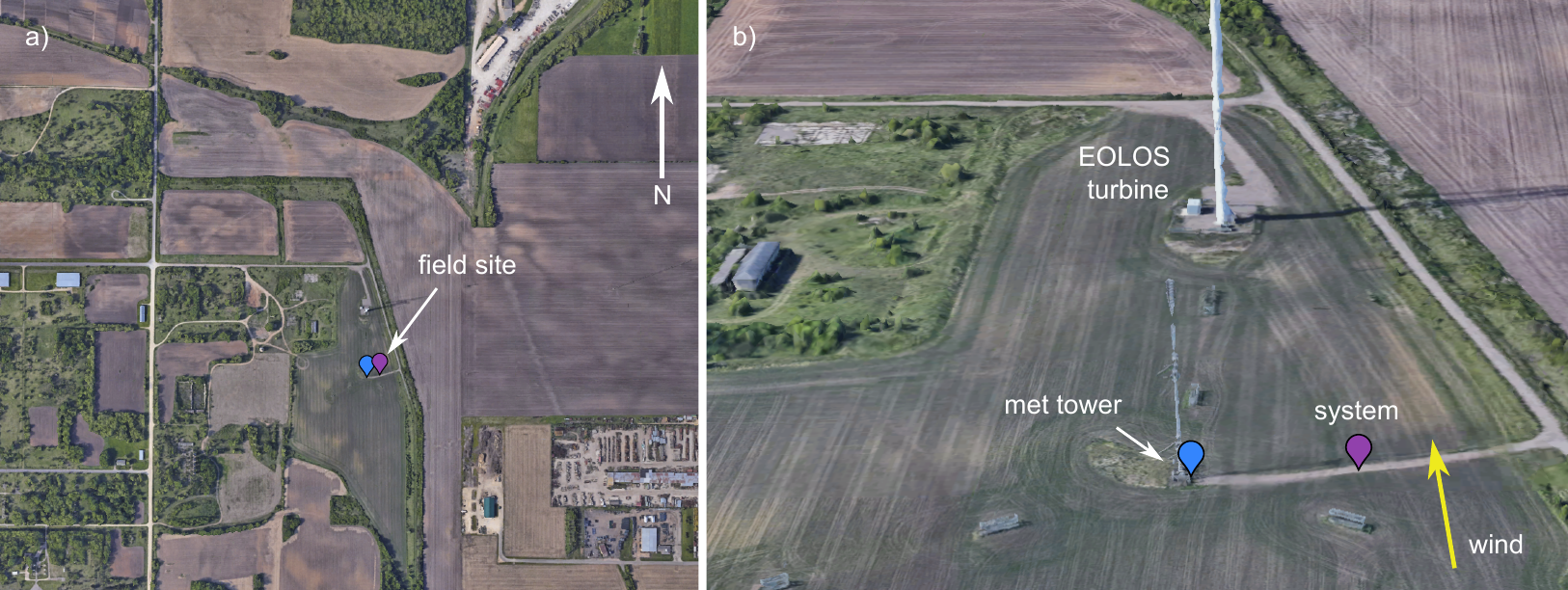}
		\caption{Satellite images of field site for experiments in Rosemount, Minnesota, indicating positions of the system deployment in April 2022 and of the meteorological tower, as well as the mean wind direction from the deployment.}
		\label{fig:exp:fieldsite}
	\end{figure}

\begin{figure}[h]
		\centering
		\includegraphics[width=1.0\textwidth]{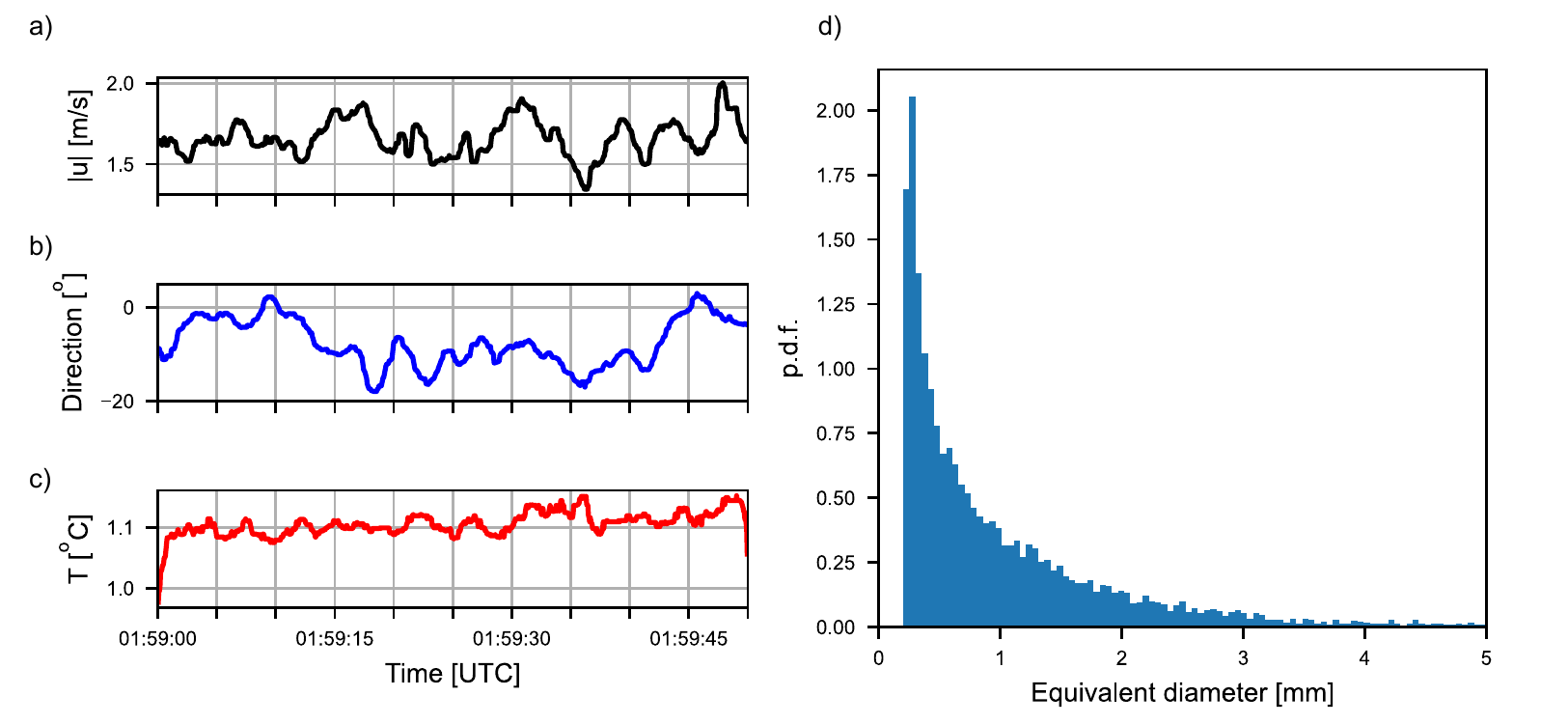}
		\caption{a--c) Meteorological tower data from the 10\,m sonic anemometer, showing a) wind speed, b) wind direction, and c) temperature. d) Snow particle equivalent diameters from the deployment period.}
		\label{fig:exp:conditions}
	\end{figure}

	A series of experiments measuring natural snowfall were undertaken during the winter season of 2021/2022. All measurements were performed at the University of Minnesota Outreach, Research and Education (UMore) Park, in Rosemount, Minnesota, near the EOLOS wind energy research station. The region surrounding the deployment location of the GAIA-PTV system is depicted in Fig.~\ref{fig:exp:fieldsite}, and consists of relatively flat farmland, apart from the meteorological (met) tower nearby. The met tower is equipped with SAT3 Campbell Scientific Sonic anemometers at elevations of 10, 30, 80, and 129\,m above ground, sampling at 20 Hz, and cup-and-vane anemometers at elevations of 7, 27, 52, 77, 102, and 126\,m above ground. Further details concerning the met tower are provided in \citep{toloui2014measurement,heisel2018spatial}. 
	The  data analyzed herein was collected on April 17, 2022 between the hours of 20:39 and 23:00 CST. Though multiple datasets were collected during this period, here we present an exemplary subset of this data from 20:59 to 21:00. Wind conditions and air temperature during this period are shown in Fig.~\ref{fig:exp:conditions}a--c, where speeds of approximately 1.7 m/s were steady within $\pm$0.3 m/s, measured from the 10\,m sonic anemometer. 
	
	
	
\begin{figure}[h]
		\centering
		\includegraphics[width=0.85\textwidth]{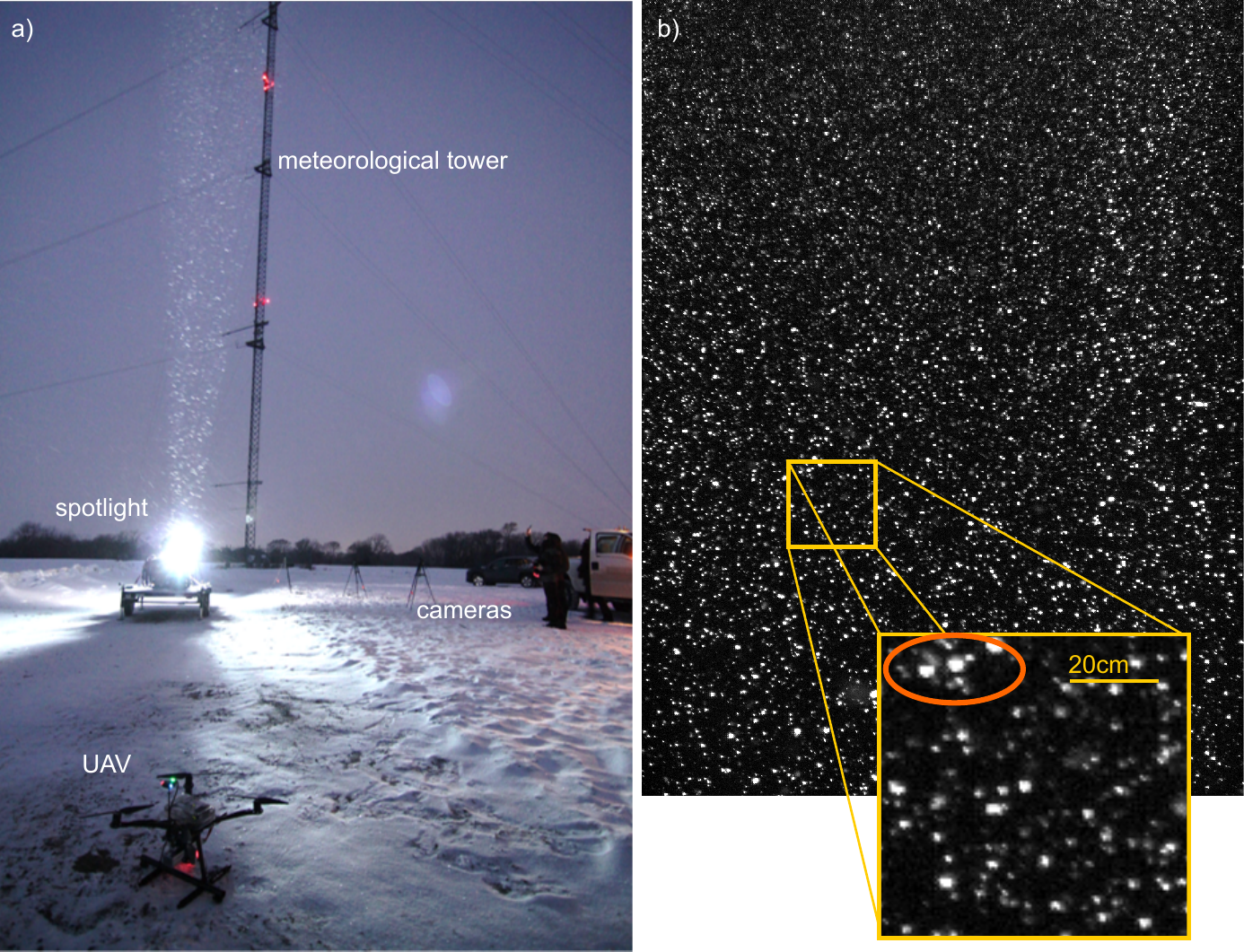}
		\caption{a) Photo of the field site indicating relevant equipment, and b) sample snow particle image.}
		\label{fig:exp:photo}
	\end{figure}

	The morphologies of the snow particles that precipitated during the deployment period were analyzed using a digital inline holography system similar to that used by \cite{Nemes2017}. The equivalent diameters of the particles are shown in Fig.~\ref{fig:exp:conditions}d, where particles typically measured less than 5\,mm in diameter. It may be noted that the p.d.f. differs here from those shown by \cite{Nemes2017} and \cite{CLi2021}, wherein a distinct peak in the distribution could be identified, whereas here the p.d.f. increases towards zero. This is due to the increased accuracy of the particle identification used herein that is able to identify snow particles over a larger range of sizes using machine learning \citep{li2022snow}.
	The field site and equipment used are depicted in Fig.~\ref{fig:exp:fieldsite}a, in which can be seen the layout of the spotlight illumination, met tower, camera system, UAV, and an example of raw particle images obtained. The layout is also schematically illustrated in the inset of Fig.~\ref{fig:methods:schematic}. Cameras were mounted on tripods in a ``fan'' array spanning 90 degrees of a circular arc, 5.5\,m in radius. All were tilted upward to be oriented at 58 degrees from horizontal, resulting in the varying magnification across the field of view in Fig.~\ref{fig:exp:photo}b. This resulted in a measurement volume of approximately $4\times4\times6$\,m in $x$, $y$, and $z$, the streamwise, spanwise, and wall-normal directions, respectively. Note that the linear size of the field of view is of the same order of magnitude of the integral scale of the flow at that distance from the ground, $\simeq k_v z \simeq 4$\,m, based on the von Karman constant $k_v$ and the mixing length assumption in turbulent boundary layers. Cameras recorded at 200 fps in decimated mode with 732$\times$968 pixels, providing spatial resolution of $\approx$6.3\,mm per pixel at the center of the FOV. For a given dataset, 10,000 frames were recorded in each camera, providing 50 second duration sequences. 
	
	\subsection{Calibration}\label{sec:experiment:calibration}
	
	A measurement volume of approximately 4\,m$\times$4\,m$\times$6\,m was calibrated (after the particle image datasets were collected) as depicted in the inset of Fig.~\ref{fig:methods:schematic}, where red and green markers indicate the reconstructed trajectory of the UAV as it flew through the measurement region. Cameras recorded the UAV lights at 30 fps during this sequence and resulted in 4063 frames in which both lights could be seen by all four cameras and successfully identified. This resulted in mean reprojection errors of 0.275 pixels across the four cameras, or approximately 1.73\,mm in the center of the measurement domain. This showed an improvement upon the validation experiment, likely due to the longer calibration sequence capturing a greater number of calibration points. Please note that 1.73\,mm is of the same order as $d_s$, mean snow particle diameter, and $<1\%$ of $I_d$, mean inter-particle distance (see \ref{sec:results:clustering}).
	
	\section{Results}\label{sec:results}
	\subsection{Particle trajectories}\label{sec:results:trajectories}
	
	Having obtained this camera calibration, snow particle images could be processed using 3D PTV. In the following, the coorindate system is defined such that $x$ is the streamwise direction, $y$ is spanwise, and $z$ is vertical. 
	
\begin{figure}[h]
		\centering
		\includegraphics[width=0.5\textwidth]{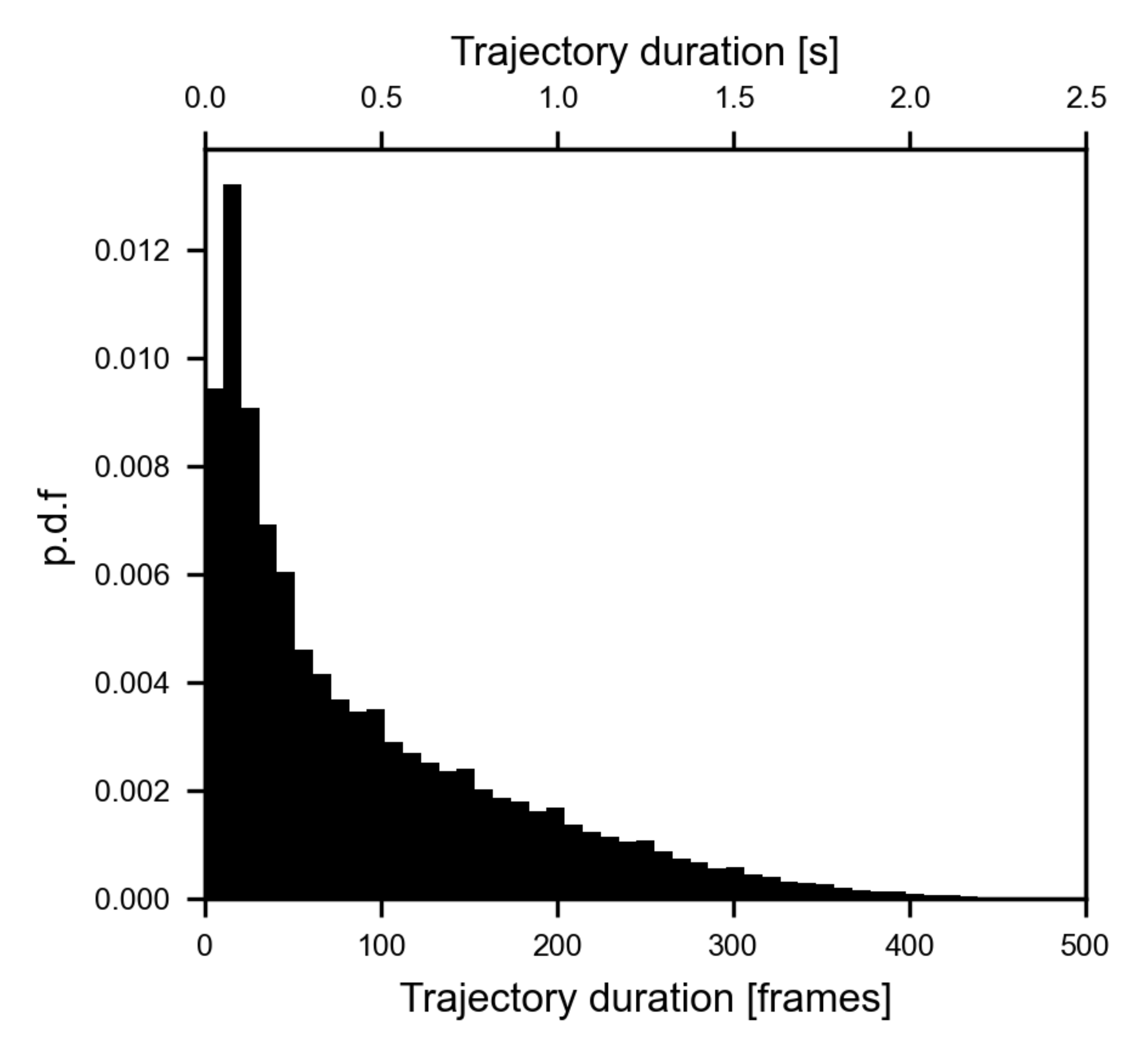}
		\caption{Probability density function of snow particle tracking durations for continuous ``long'' trajectories.}
		\label{fig:results:duration}
	\end{figure}

	The distribution of snow particle trajectory durations, obtained from 250,619 particles tracked over 10,000 frames, is shown in Fig.~\ref{fig:results:duration}. Due to the challenge in tracking particle motions in 3D, durations are skewed towards lower values, and taper off at the long end by approximately 400 frames, or 2 seconds. This does not include particle trajectories shorter than 5 frames, which are not considered valid trajectories for studying particle kinematics as they have not passed the initialization phase and developed sufficient duration. Approximately 1800 ``long'' particle trajectories ($>4$ frames) and 7000 ``short'' trajectories are obtained at each frame. These longer trajectories are continuously tracked until they either exit the domain or become ``inactive'' (i.e., lost). The number of snow particles present in the measurement volume, on average, is $\approx11,000$, based on the number of particles identified in all four camera frames. Thus the measurement yield of reconstructed particle positions, out of all particles physically in the measurement domain, is estimated as 80\% while the yield of long trajectories is be approximated as 20\%. This lower estimate compared to the confetti validation experiment is most likely due to imperfect view overlap from different cameras at the edges of the FOV and poorer image quality from the upper region of the measurement volume.
	
\begin{figure}[h]
		\centering
		\includegraphics[width=1.0\textwidth]{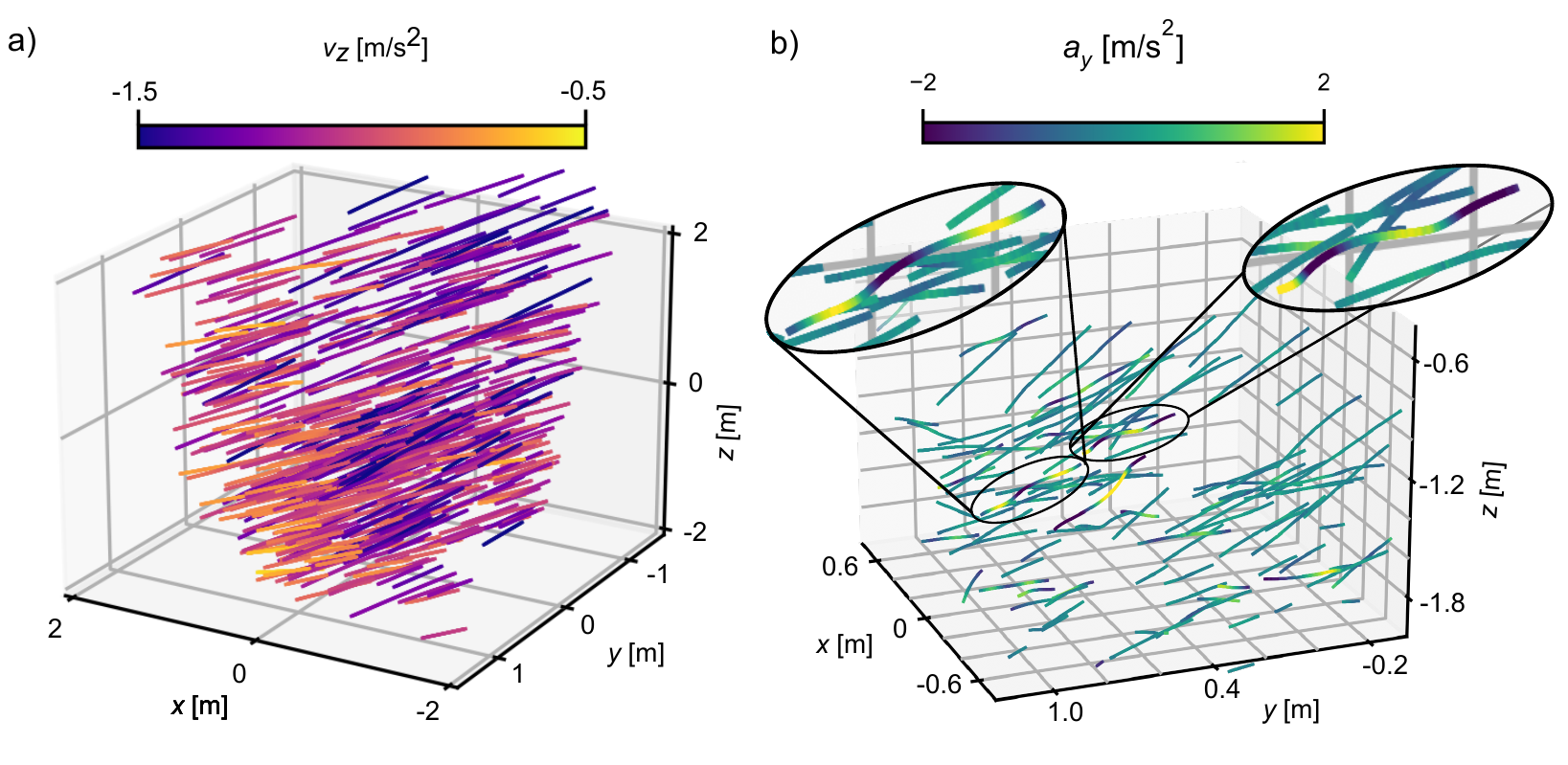}
		\caption{Sample snow particle trajectories plotted at different viewing angles and magnification, colored by spanwise acceleration, $a_y$. The $x$-axis is the mean streamwise direction, and the coordinate system origin is at the center of the measurement volume.}
		\label{fig:results:sampletracks}
	\end{figure}

	A sample set of snow particle trajectories is shown in Fig.~\ref{fig:results:sampletracks} from two different vantage points. As can be seen in Fig.~\ref{fig:results:sampletracks}a, when viewed from the side particle trajectories appear relatively linear, elongated in the streamwise direction. This type of view is similar to what would be captured using a 2D imaging method. Closer inspection of Fig.~\ref{fig:results:sampletracks}c reveals more complex kinematics  of the particles, with subtle oscillations in their curvature. This is emphasized through the colormap applied to the trajectories. Here, the lateral acceleration component, $a_y$, is chosen as it particularly emphasizes the three-dimensionality of the particle motion that cannot be captured with planar or 2D imaging methods. The particles along these trajectories accelerate back and forth, suggesting non-negligible interactions with the turbulent flow and the different scales of vortical motions \citep{JLi2021}. Furthermore, this behavior is not observable for all neighboring particles; some are observed to translate relatively linearly along the primary flow direction as they settle. This could be speculated to be the result of differences in the size and morphology of the snow particles, affecting their inertial properties and their ability to follow different scales of the flow, despite different particles sampling similar regions in the flow. Note that the mean inter-particle distance is $\simeq 0.18$\,m, or $\sim2\lambda_T$, where $\lambda_T$ is the Taylor microscale. Estimated by the nearby sonic anemometer, $\lambda_T$ provides reasonable estimates of the thickness of the shear layers and the size of the vortex cores \citep{Heisel2021}, implying that nearby snow particles may still sample different flow topologies.
	
	\subsection{Acceleration and curvature}
	
\begin{figure}[h]
		\centering
		\includegraphics[width=\textwidth]{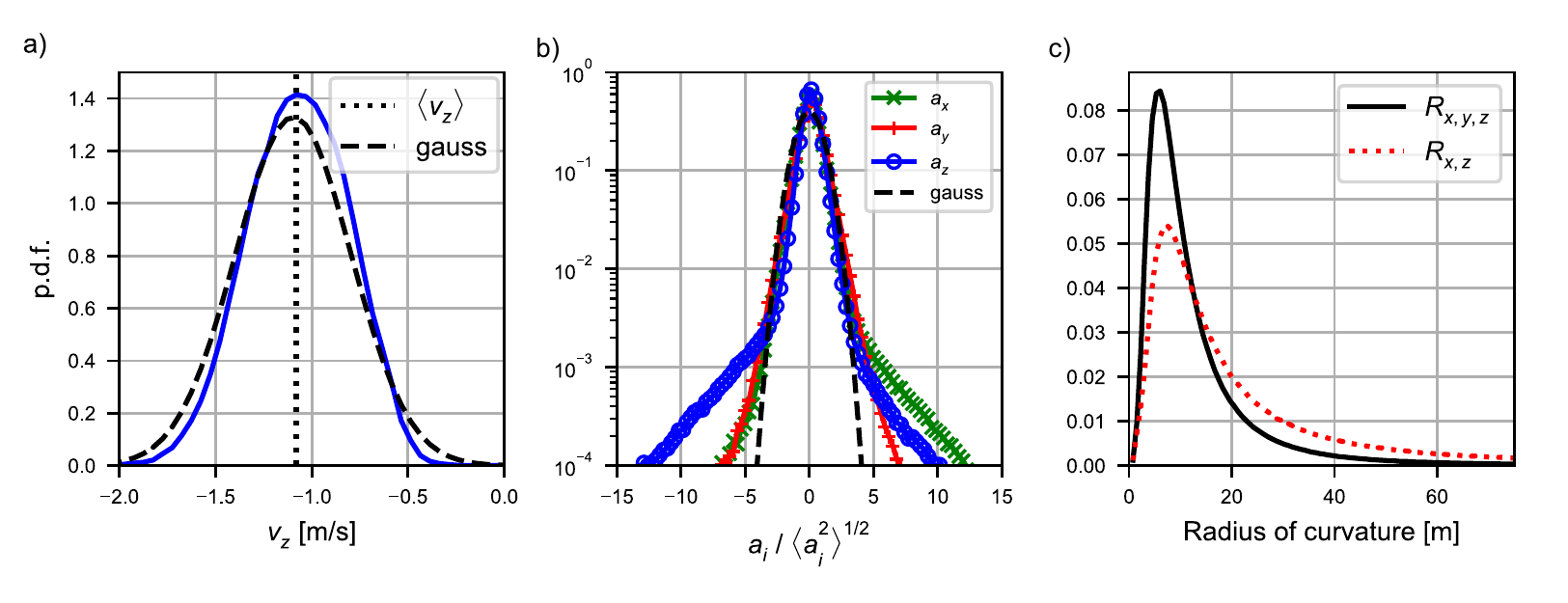}
		\caption{Probability density functions of a) vertical settling velocity, b) acceleration components, and c) radii of curvature. In a) and b), the dashed lines indicate Gaussian distributions for comparison.)}
		\label{fig:results:curv-accel}
	\end{figure}

	The statistics of the particle kinematics, explored qualitatively in the previous section, are presented in Fig.~\ref{fig:results:curv-accel} plotting p.d.f.s of settling velocity, acceleration components, and curvature. Settling velocity displays an approximately normal distribution, weakly skewed towards lower fall speeds. In terms of acceleration, the spanwise component, $a_y$, is generally of similar magnitude to $a_x$ and $a_z$, though their distribution tails differ, and as such these spanwise motions are important to capture. All three components display wider tails compared to a normal distribution, where $a_z$ in particular is skewed towards negative accelerations. The streamwise component, $a_x$, on the other hand, is more strongly skewed with positive accelerations.
	
	Curvature is given as
	\begin{equation}
		\rho=\frac{1}{R}=\frac{\|\bf{v} \times a\|}{\|\bf{v}\|^3}
	\end{equation}
	where $\rho$ is curvature, $R$ is the radius of curvature, $\bf{v}$ and $\bf{a}$ are velocity and acceleration vectors, respectively, and double brackets indicate taking the L2-norm. The distribution of $R$, plotted in Fig.~\ref{fig:results:curv-accel}c, is calculated in both 2D and 3D, where in the 2D case only the $x$ and $z$ components of $\bf{v}$ and $\bf{a}$ are used, simulating the components that would be resolved if a planar imaging system were used that was appropriately aligned parallel to the streamwise direction. Markedly different distributions are shown between these two, where the 2D case is skewed towards larger weaker curvature, with peaks in their distributions of approximately 8\,m and 6\,m for the 2D and 3D cases, respectively. This agrees with the observed behavior in the sample trajectories from Fig.~\ref{fig:results:sampletracks}c, which showed significant lateral components in acceleration and should be expected given the probability density functions of acceleration components shown in Fig.~\ref{fig:results:curv-accel}b. During parts of trajectories where the particle may oscillate in the $y$-direction, a 2D tracking system would not register this out-of-plane motion and thus return higher radii of curvature for such tracks. Please note that particle inertial properties are known to affect the curvature and acceleration distribution \cite[see e.g.,][]{bec2006acceleration}, as heavier-than-fluid particles cannot be trapped in vortical flows. Hence, it can be expected that the peak of curvature should be larger than the expected size of vortex cores. 
	
	\subsection{Clustering}\label{sec:results:clustering}
	
\begin{figure}[h]
		\centering
		\includegraphics[width=1.0\textwidth]{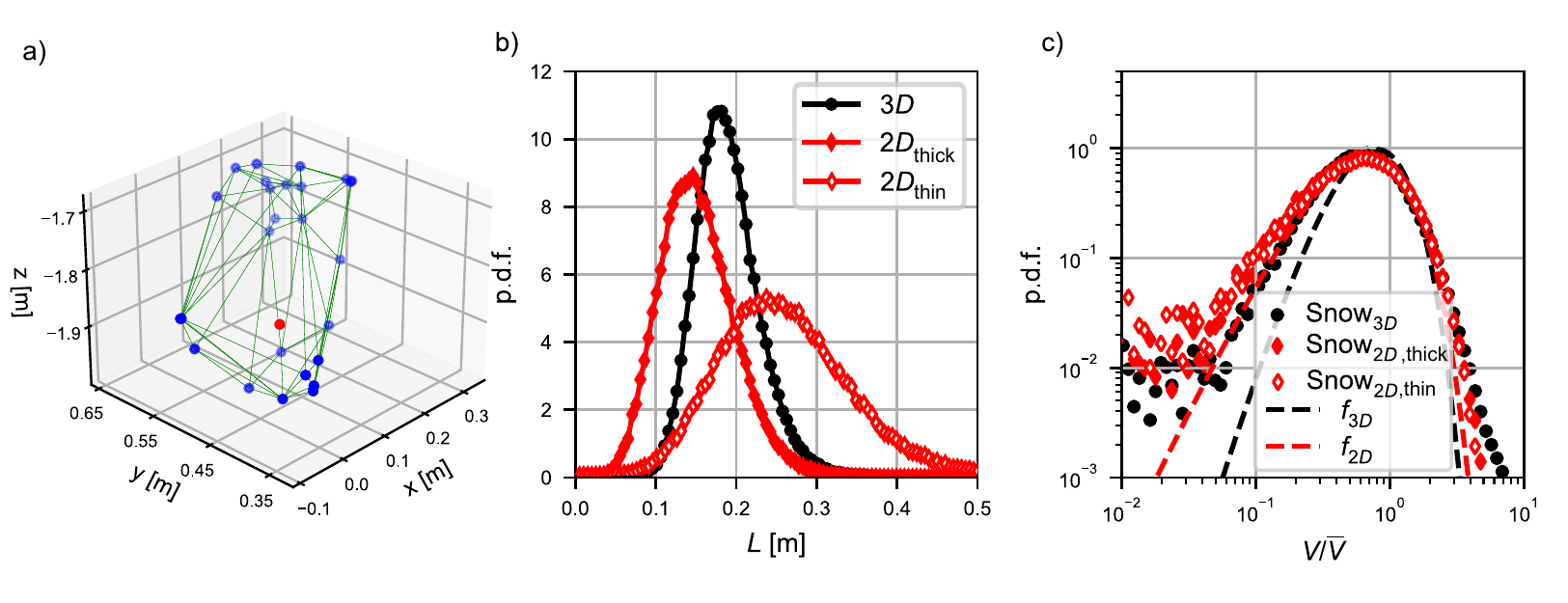}
		\caption{a) Example of a single Vorono\"{i} cell produced by the tessellation, where the red dot indicates the snow particle, blue dots indicate vertices, and green lines are the ridges of the cell; b) Probability density functions of cell length scales, $L$, comparing the 2D and 3D cases; c) Probability density functions in log--log space of cell volumes, compared against models from \cite{ferenc2007size} for randomly distributed particles given by Equations \ref{eqn:f_2D} and \ref{eqn:f_3D}.}
		\label{fig:results:clustering}
	\end{figure}

	Three-dimensional particle cloud reconstructions are particularly advantageous, as compared to 2D planar imaging, because they also enable direct quantification of particle concentration statistics and clustering behaviors without any 2D assumptions. This can be achieved using Vorono\"{i} diagrams, spatial tessellations of the particle cloud, wherein the volume of individual cells produced by the tessellation are inversely proportional to the local concentration. This method is preferable to other approaches, such as box-counting, as it avoids biasing due to input parameter choices such as bin size, and is also robust to particle sub-sampling (up to 50\%) effects often unavoidable in experimental data \citep{monchaux2012measuring}. The shape and extent of a Vorono\"{i} cell are defined such that all points within the cell (except the edges) are closer to the single particle lying within the cell than they are to any neighboring particles outside the cell. Thus, each Vorono\"{i} cell is associated with a given particle, and the vertices and borderlines themselves are equidistant to the neighboring particles. An example of a single Vorono\"{i} cell is shown in Figure~\ref{fig:results:clustering}a, showing the complex topology of the cell in 3D space, while the p.d.f. of Vorono\"{i} cell size is shown in Fig.~\ref{fig:results:clustering}b. Here, the length scales associated with the cell size for both the full 3D data as well as quasi-2D data is shown. The 2D case is simulated for comparison by computing the 2D Vorono\"{i} tessellation on particles found within sub-volumes oriented in the $x$--$z$ plane, with their y-dimension removed. Two different sub-volume thicknesses are compared, of 0.3\,m and 0.1\,m, similar to that used in previous 2D planar snow tracking experiments \citep{Nemes2017,CLi2021,JLi2021}. In all cases, 2D or 3D the data are taken from particles within $\pm1$\,m of the measurement domain center (in all directions) in order to avoid any unwanted effects near the boundaries. The length scale, $L$, for the 3D case is the cube-root of the cell volume, while for the 2D case $L$ is the square-root of the cell area. The distributions of $L$ peak at 0.18\,m for the 3D, indicative of true inter-particle spacing, $I_d$, whereas the 2D cases are affected by the sub-volume thickness. The thicker 2D sub-volume results in smaller $L$, due to the fact that more particles are projected onto the $x$--$z$ plane, compared to the thin 2D sub-volume \cite{monchaux2012measuring}.
	
	Normalizing the volumes (or areas for the 2D case) by their respective means, these data can be compared to randomly distributed particles (Fig.~\ref{fig:results:clustering}c), which, following \cite{ferenc2007size}, can be described for the 2D and 3D cases, respectively, as
	\begin{equation}
		f_{2 D}(x)=\frac{343}{15} \sqrt{\frac{7}{2 \pi}} x^{5 / 2} e^{-7 x / 2} 
		\label{eqn:f_2D}
	\end{equation}
	\begin{equation}
		f_{3 D}(x)=\frac{3125}{24} x^4 e^{-5 x}
		\label{eqn:f_3D}
	\end{equation}
	
	Deviations from these distributions indicate clustering of particles due to flow and particle inertia. In all comparisons between snow data and profiles of $f_{2D}$ and $f_{3D}$, it can be seen that the snow particles display  clustering behavior. Evidence of such clustering can be qualitatively observed in the raw images, as indicated in the inset of Fig.~\ref{fig:exp:photo}b, though it must be remembered that the image is a projection of all particles across the depth of the volume and therefore the scale from the images may not be indicative of true spacing. Furthermore, the results of the distributions suggest that 2D measurement, of the same snow particles, would have slightly underestimated the extent of this clustering behavior. This is shown by the disparity between the distribution for the 3D experimental data and the profile for $f_{3D}$, which is greater than that of the 2D data. This is true for both the ``thick'' or ``thin'' 2D data, where it can also be noted that the normalization here largely removes the differences observed in Fig.~\ref{fig:results:clustering}b. Our results are largely consistent with the results shown by \cite{monchaux2012measuring} concerning 2D/3D biases in Vorono\"{i} analysis, wherein differences between 2D and 3D treatments are minor though appreciable. It may be speculated, however, that the extent to which these differences between 2D or 3D treatments are manifested may be influenced by the particular type of flow (e.g., a turbulent boundary layer) and the statistically persistent 3D vortical structures it contains.
	
	\section{Conclusions and Discussion}
	
	The GAIA-PTV system presented herein has been demonstrated to provide unprecedented Lagrangian tracking of snow settling motions in 3D using a multiview imaging array and UAV-based calibration approach. Snow particles are tracked within a volume measuring approximately 4\,m$\times$4\,m$\times$6\,m centered at a height 10\,m above ground providing the ability to capture long trajectories that free from flow disturbances due to the ground. The results presented herein provide a snapshot of the analyses that are enabled by GAIA-PTV, and demonstrate the utility of measuring particle kinematics in 3D over 2D. 
	
	Importantly, this work enables further systematic investigations of snow settling dynamics. With the measurement system demonstrated, experiments can be conducted across a range of atmospheric turbulence conditions to offer unique insights into particle kinematics and clustering. Furthermore, the GAIA-PTV system can be extended to measuring transport of other particles in atmospheric flows, such as droplet sprays and dispersed pollen. The system was designed to be both robust in the field, as has been tested under the harsh winter conditions, and scalable depending on the needs of the flow phenomena of interest. 
	
	In future work, two key limitations of the current system could be improved upon. Firstly, as with most 3D PTV systems, the Lagrangian tracking capability is limited in part by the domain size and the fact that the domain is fixed. If the current system could be made mobile, such as by using UAV-mounted cameras instead of fixed tripod-mounted cameras on the ground, it would enable much longer tracking times to study particle dispersion, for example. In a similar vein as has been performed with translating laboratory imaging systems (e.g., the ``flying PIV'' by \cite{zheng2013flying}) the system could follow along with the particles of interest, keeping them within the measurement domain. Such advances are challenging, however, primarily due to the problem of calibrating this type of system. 
	
	Second, the need for artificial illumination as used herein constrains the measurement opportunities to the hours of darkness, when the particles of interest can be illuminated against the night sky. Advances in particle detection methods capable of dealing with natural daylight illumination could significantly increase use-cases, as well as simplify the deployment requirements without the need for a high-wattage light sources. The primary challenge is in overcoming the low contrast available when imaging during daylight, but rapid developments in the field of computer vision and object detection may yield promising tools for improvements in this regard.
	
	\bmhead{Acknowledgments}
	
	This work was supported by the National Science Foundation under grants CBET-2018658 and AGS-1822192.
	
	\section*{Declarations}
	
	\begin{itemize}
		
		\item Ethical approval: Not applicable.
		
		\item Author contributions: J.H. and M.G. oversaw the study conception and design. N.B. wrote the manuscript text, generated figures, and was the primary contributor to the measurement system development. P.H. contributed to the development of the measurement system. All authors contributed to the field experiments.
		
		\item Funding: This work was supported by the National Science Foundation under grants CBET-2018658 and AGS-1822192.
		
		\item Availability of data and materials: The data that support the findings of this study are available from the corresponding author, J.H., upon reasonable request.
		
		\item Competing interests: The authors declare no conflicts of interest.
		
	\end{itemize}


\begin{thebibliography}{34}
		\providecommand{\natexlab}[1]{#1}
		\providecommand{\url}[1]{{#1}}
		\providecommand{\urlprefix}{URL }
		\providecommand{\doi}[1]{\url{https://doi.org/#1}}
		\providecommand{\eprint}[2][]{\url{#2}}
		\bibcommenthead
		
		\bibitem[{Abraham and Hong(2020)}]{abraham2020dynamic}
		Abraham A, Hong J (2020) Dynamic wake modulation induced by utility-scale wind
		turbine operation. Applied Energy 257:114,003
		
		\bibitem[{Bec et~al(2006)Bec, Biferale, Boffetta, Celani, Cencini, Lanotte,
			Musacchio, and Toschi}]{bec2006acceleration}
		Bec J, Biferale L, Boffetta G, et~al (2006) Acceleration statistics of heavy
		particles in turbulence. Journal of Fluid Mechanics 550:349--358
		
		\bibitem[{Discetti and Coletti(2018)}]{discetti2018volumetric}
		Discetti S, Coletti F (2018) Volumetric velocimetry for fluid flows.
		Measurement Science and Technology 29(4):042,001
		
		\bibitem[{Emanuel(2018)}]{emanuel2018100}
		Emanuel K (2018) 100 years of progress in tropical cyclone research.
		Meteorological Monographs 59:15--1
		
		\bibitem[{Ferenc and N{\'e}da(2007)}]{ferenc2007size}
		Ferenc JS, N{\'e}da Z (2007) On the size distribution of poisson voronoi cells.
		Physica A: Statistical Mechanics and its Applications 385(2):518--526
		
		\bibitem[{Garrett and Yuter(2014)}]{garrett2014observed}
		Garrett TJ, Yuter SE (2014) Observed influence of riming, temperature, and
		turbulence on the fallspeed of solid precipitation. Geophysical Research
		Letters 41(18):6515--6522
		
		\bibitem[{Hartley and Zisserman(2003)}]{hartley2003multiple}
		Hartley R, Zisserman A (2003) Multiple view geometry in computer vision.
		Cambridge University Press
		
		\bibitem[{Heisel et~al(2018)Heisel, Dasari, Liu, Hong, Coletti, and
			Guala}]{heisel2018spatial}
		Heisel M, Dasari T, Liu Y, et~al (2018) The spatial structure of the
		logarithmic region in very-high-reynolds-number rough wall turbulent boundary
		layers. Journal of Fluid Mechanics 857:704--747
		
		\bibitem[{Heisel et~al(2021)Heisel, de~Silva, Hutchins, Marusic, and
			Guala}]{Heisel2021}
		Heisel M, de~Silva C, Hutchins N, et~al (2021) {Prograde vortices, internal
			shear layers and the Taylor microscale in high-Reynolds-number turbulent
			boundary layers}. Journal of Fluid Mechanics 920:715--731
		
		\bibitem[{Hommema and Adrian(2003)}]{hommema2003packet}
		Hommema SE, Adrian RJ (2003) Packet structure of surface eddies in the
		atmospheric boundary layer. Boundary-Layer Meteorology 106(1):147--170
		
		\bibitem[{Hong et~al(2014)Hong, Toloui, Chamorro, Guala, Howard, Riley, Tucker,
			and Sotiropoulos}]{hong2014natural}
		Hong J, Toloui M, Chamorro LP, et~al (2014) {Natural snowfall reveals
			large-scale flow structures in the wake of a 2.5-MW wind turbine}. Nature
		communications 5(1):1--9
		
		\bibitem[{Huang et~al(2015)Huang, Liu, Chen, and Nasiri}]{huang2015detection}
		Huang J, Liu J, Chen B, et~al (2015) Detection of anthropogenic dust using
		calipso lidar measurements. Atmospheric Chemistry and Physics
		15(20):11,653--11,665
		
		\bibitem[{Kok et~al(2018)Kok, Ward, Mahowald, and Evan}]{kok2018global}
		Kok JF, Ward DS, Mahowald NM, et~al (2018) Global and regional importance of
		the direct dust-climate feedback. Nature Communications 9(1):1--11
		
		\bibitem[{Lapotre et~al(2016)Lapotre, Ewing, Lamb, Fischer, Grotzinger, Rubin,
			Lewis, Ballard, Day, Gupta et~al}]{lapotre2016large}
		Lapotre M, Ewing R, Lamb M, et~al (2016) Large wind ripples on mars: A record
		of atmospheric evolution. Science 353(6294):55--58
		
		\bibitem[{Li et~al(2021{\natexlab{a}})Li, Lim, Berk, Abraham, Heisel, Guala,
			Coletti, and Hong}]{CLi2021}
		Li C, Lim K, Berk T, et~al (2021{\natexlab{a}}) Settling and clustering of snow
		particles in atmospheric turbulence. Journal of Fluid Mechanics 912
		
		\bibitem[{Li et~al(2021{\natexlab{b}})Li, Abraham, Guala, and Hong}]{JLi2021}
		Li J, Abraham A, Guala M, et~al (2021{\natexlab{b}}) Evidence of preferential
		sweeping during snow settling in atmospheric turbulence. Journal of Fluid
		Mechanics 928
		
		\bibitem[{Li et~al(2022)Li, Guala, and Hong}]{li2022snow}
		Li J, Guala M, Hong J (2022) Snow particle analyzer for simultaneous
		measurements of snow density and morphology. arXiv preprint arXiv:220911129
		
		\bibitem[{Lourakis and Argyros(2009)}]{lourakis2009sba}
		Lourakis MI, Argyros AA (2009) Sba: A software package for generic sparse
		bundle adjustment. ACM Transactions on Mathematical Software (TOMS)
		36(1):1--30
		
		\bibitem[{Lundquist et~al(2017)Lundquist, Wilczak, Ashton, Bianco, Brewer,
			Choukulkar, Clifton, Debnath, Delgado, Friedrich
			et~al}]{lundquist2017assessing}
		Lundquist JK, Wilczak JM, Ashton R, et~al (2017) {Assessing state-of-the-art
			capabilities for probing the atmospheric boundary layer: the XPIA field
			campaign}. Bulletin of the American Meteorological Society 98(2):289--314
		
		\bibitem[{Monchaux(2012)}]{monchaux2012measuring}
		Monchaux R (2012) Measuring concentration with vorono{\"\i} diagrams: the study
		of possible biases. New Journal of Physics 14(9):095,013
		
		\bibitem[{Morris et~al(2007)Morris, Stolpa, Slaboch, and
			Klewicki}]{morris2007near}
		Morris SC, Stolpa SR, Slaboch PE, et~al (2007) Near-surface particle image
		velocimetry measurements in a transitionally rough-wall atmospheric boundary
		layer. Journal of Fluid Mechanics 580:319--338
		
		\bibitem[{Muller et~al(2020)Muller, Hemelrijk, Westerweel, and
			Tam}]{muller2020calibration}
		Muller K, Hemelrijk C, Westerweel J, et~al (2020) Calibration of multiple
		cameras for large-scale experiments using a freely moving calibration target.
		{Experiments in Fluids} 61(1):1--12
		
		\bibitem[{Nemes et~al(2017)Nemes, Dasari, Hong, Guala, and Coletti}]{Nemes2017}
		Nemes A, Dasari T, Hong J, et~al (2017) Snowflakes in the atmospheric surface
		layer: observation of particle–turbulence dynamics. Journal of Fluid
		Mechanics 814:592--613
		
		\bibitem[{Novara et~al(2019)Novara, Schanz, Geisler, Gesemann, Voss, and
			Schr{\"o}der}]{novara2019multi}
		Novara M, Schanz D, Geisler R, et~al (2019) Multi-exposed recordings for 3d
		lagrangian particle tracking with multi-pulse shake-the-box. {Experiments in
			Fluids} 60(3):1--19
		
		\bibitem[{Perks et~al(2020)Perks, Dal~Sasso, Hauet, Jamieson, Le~Coz, Pearce,
			Pe{\~n}a-Haro, Pizarro, Strelnikova, Tauro et~al}]{perks2020towards}
		Perks MT, Dal~Sasso SF, Hauet A, et~al (2020) Towards harmonisation of image
		velocimetry techniques for river surface velocity observations. Earth System
		Science Data 12(3):1545--1559
		
		\bibitem[{Schanz et~al(2016{\natexlab{a}})Schanz, Gesemann, and
			Schr{\"o}der}]{schanz2016shake}
		Schanz D, Gesemann S, Schr{\"o}der A (2016{\natexlab{a}}) Shake-The-Box:
		Lagrangian particle tracking at high particle image densities. {Experiments
			in Fluids} 57(5):1--27
		
		\bibitem[{Schanz et~al(2016{\natexlab{b}})Schanz, Huhn, Gesemann, Dierksheide,
			van~de Meerendonk, Manovski, and Schr{\"o}der}]{schanz2016towards}
		Schanz D, Huhn F, Gesemann S, et~al (2016{\natexlab{b}}) {Towards
			high-resolution 3D flow field measurements at cubic meter scales}. In:
		Proceedings of the 18th International Symposium on the Application of Laser
		and Imaging Techniques to Fluid Mechanics, Springer
		
		\bibitem[{Tan et~al(2020)Tan, Salibindla, Masuk, and Ni}]{tan2020introducing}
		Tan S, Salibindla A, Masuk AUM, et~al (2020) Introducing openlpt: new method of
		removing ghost particles and high-concentration particle shadow tracking.
		{Experiments in Fluids} 61(2):1--16
		
		\bibitem[{Terra et~al(2019)Terra, Sciacchitano, and
			Shah}]{terra2019aerodynamic}
		Terra W, Sciacchitano A, Shah Y (2019) Aerodynamic drag determination of a
		full-scale cyclist mannequin from large-scale ptv measurements. {Experiments
			in Fluids} 60(2):1--11
		
		\bibitem[{Theriault et~al(2014)Theriault, Fuller, Jackson, Bluhm, Evangelista,
			Wu, Betke, and Hedrick}]{theriault2014protocol}
		Theriault DH, Fuller NW, Jackson BE, et~al (2014) A protocol and calibration
		method for accurate multi-camera field videography. Journal of Experimental
		Biology 217(11):1843--1848
		
		\bibitem[{Toloui et~al(2014)Toloui, Riley, Hong, Howard, Chamorro, Guala, and
			Tucker}]{toloui2014measurement}
		Toloui M, Riley S, Hong J, et~al (2014) Measurement of atmospheric boundary
		layer based on super-large-scale particle image velocimetry using natural
		snowfall. {Experiments in Fluids} 55(5):1--14
		
		\bibitem[{Wei et~al(2021)Wei, Brownstein, Cardona, Howland, and
			Dabiri}]{wei2021near}
		Wei NJ, Brownstein ID, Cardona JL, et~al (2021) Near-wake structure of
		full-scale vertical-axis wind turbines. Journal of Fluid Mechanics 914
		
		\bibitem[{Wieneke(2012)}]{wieneke2012iterative}
		Wieneke B (2012) Iterative reconstruction of volumetric particle distribution.
		Measurement Science and Technology 24(2):024,008
		
		\bibitem[{Zheng and Longmire(2013)}]{zheng2013flying}
		Zheng S, Longmire EK (2013) {Flying PIV investigation of vortex packet
			evolution in perturbed boundary layers}. In: PIV13; 10th International
		Symposium on Particle Image Velocimetry, Delft, The Netherlands, July 1-3,
		2013, Citeseer
		
	\end{thebibliography}

\end{document}